# REMOTE MONITORING L.V. SWITCHBOARD OWER STATUS WITH 5G NEW RADIO NETWORK APPLICATION

Tony Tsang and Lam Sai Chung

Centre of International Education, Hong Kong College of Technology,
Hong Kong.

*ABSTRACT*

*In a power distribution system of most of the buildings, A Low Voltage (L.V.) switchboard was applied to protect the system. There are different components in the switchboard (e.g. circuit breakers, over current protection relay, earth fault protection relay etc.). There are also some components for measure the power quality which is the power analyser. A power analyser (including Voltmeter, Ammeter, Multi-meter) is using to measure the power quality in the electrical power distribution system. Most of the electrical power distributions systems have been connected the power analysers to a building management system. The power analysers connected to a computer in a fixed position that means if someone wants to check the power distribution status, the person needs to go to the computer to check. In this project, a system would be made to monitoring the power status by 5G New Radio (NR) Network application with android smartphone. Since 5G NR Network can provide a theoretical peak download capacity of 20 Gigabits per second. It would be more convenient to monitoring the power status in different place by checking though in the Internet of Things (IoT)*
.

## 1. INTRODUCTION

The power analyser would be used to measure the power quality in an electrical power system. Most of the power analysers are in digital nowadays. For the digital power analyser, it can show the voltage, ampere, power factor, Total Harmonic Distortion etc., which have more function with the analogue power meter. For recording the information, the power analysers would be connected to the building management system (BMS) by the supervisory control and data acquisition (SCADA).

There are different methods to connect to different PLC systems, one of the methods is connecting by RS485 Modbus. The RS in the RS485 means Recommended Standard, which also as known as ANSI/TIA/EIA-485-A-1998, set by EIA. The RS485 will use in the master-slave mode and Alternative master-slave mode. The master-slave mode is the master node sending command to the slave device so the slave device will do what the master node has command. The alternative master-slave mode is means different device can become the master node. The devices can send data to the other device after the device received the data, [1].

After sending the data to the database by RS485 Modbus, the data would be sent to the database on the internet and the smartphone application would be developed for accessing to the database and monitoring the power status from an android smart phone.

The aims of this project are going to design an android application to monitoring the power status by the power analyser from the L.V. switchboard. It can make more convenience to the engineer for monitoring the power supply of a building where is far from the building.

In this project, the power analyser would be installed to an electrical load for measuring the power supply's quality. The information would be sent to a database for the record. An android smartphone can access to the database by the wireless networking and shows the power quality data in the real time by an Android application.
Objectives:

- To study the power monitoring system in L.V. switchboard
- To install a power analyser to electrical load
- To study on the communication systems
- To develop a LabView graphic programme
- To build a database for recording power status
- To develop an Android application

## 2. LITERATURE REVIEW

There are different books and articles are introducing the power monitoring network in an electrical distribution system. There are also some technologies included that shows how the monitoring system works. The develop skills also be researched for the software developing.

### 2.1. Power System Monitoring and Control

"Power system monitoring and control (PSMC) is an important issue in modern electric power system design and operation" said by Hassan Bevrani, Masayuki Watanabe & Yasunori Mitani [2]. In a distribution L.V. switchboard, the power analyser usually measuring the voltage, current, power factor, Total Harmonic Distortion etc. To transmit and record the information, supervisory control and data acquisition (SCADA) system would be applied. According the supply rules of The Hongkong Electric [3] and CLP Power [4], The electricity would be applied using alternating current (A.C.) system at 50 Hz, in 380/220V (3 phase 4 wire) in Hong Kong's building normally. The Hong Kong electric also can supply 11000V for some special buildings. To measure the power quality with SCADA, the power analysers need to fulfil the power supply range and contain the signal output to the SCADA.

### 2.2. L.V. Switchboard

The L.V. switchboards are common for power distribution to the building. Different buildings have different loads. Therefore, different switchboard would be made with different size and different switchgears. The L.V. switchboards are only meet the basic requirement in the past. But nowadays, more techniques are using in the switchboard such as power analysers Francesco Muzi, Flavio D'Innocenzo [5]. The L.V. switchboard is used to design with different standards, such as IEC (International Electrotechnical Commission) and UL (Underwriters Laboratories) (Manan Deb, Tuhina Singh [6]). IEC standard would be applied for designing L.V. switchboard in Hong Kong. Usually, the digital power analysers are installed in the L.V. switchboard for the power status monitoring nowadays. To analyse the data of the power status, the power monitoring system are needed for the L.V. switchboard.

## 2.3. Fundamentals of 5G New Radio Mobile Network

Nowadays, the information technology is around our city. The network can share the information to different people by an electronic device such as mobile phone. For the power monitoring system, the information would be shared the power information to an electronic device though the network. For the internet networking, there are different types of generation. 4G networks, which is the fourth-generation macro-cell-based mobile wireless networking, is common using in the world and the 5G network is more powerful than the 4G network. The 5G New Radio (NR) network also will be common in the future for transmitting information data. In the 5G technology, it is not only the mobile and wireless piece, but also includes the wide area coverage network, said by Jonathan Rodriguez [7]. Therefore, the power monitoring system is also can use the 5G network to monitor the power quality in a faster and in higher accuracy.

## 2.4. Database Systems

There are different databases in the network, such as MySQL, Microsoft SQL and Firebase etc. The databases are on the cloud and the user can access to the database though the internet. Structured Query Language (SQL) is a language for most of the database, the database needs a SQL format code to request the database management system to access the database to take data.

The SQL are standard by the American National Standards Institute (ANSI) and the International Standards Organization (ISO) in 1986. (Weinberg, Paul N., [8]) The Database would storage the data as a table with rows and columns. The User can insert, delete, summarize or protect the data in the database with the SQL.

To apply the MySQL database, the PHP files are need. "The database can insert net data, update existing information or remove information by the php programmes." Said by Janet Valade [9]. To manage the database, phpMyAdmin would be used for setting the database such as setting the security code, manage the data and export the data etc. The database would export a JSON file to list the data, php file can help to collate the data.

## 2.5. Android Application

Android is an operation system for most of the smartphone device. The development of the android application can be done by everyone with the computer. Kurniawan Bydi [10] said that, developer can easy to write, test and deploy an application with a comprehensive set of APIs. By using the JAVA programming language, the application would be developed for the Android device. Android Studio is one of the platforms for the developer to develop an Android application. It can be installed in Windows and using the JAVA Development Kit for the development. The application can be built to debug without an Android device because there is a simulator for the debugging.

## 2.6. Power Monitoring Using Internet of Thing

In the "Industrial Greenhouse Electrical Power Monitoring Using Secure Internet-of-Things (IoT) Platform" written by Peteris Apse-Apsitis, Ansis Avotins and Ricards Porins [11] are using an

International Journal of Computer Science & Information Technology (IJCSIT) Vol 11, No 4, August 2019

electric chip to send the data to the internet by Wi-fi for the power monitoring. The HTTPS would be used and the cloud server would be connected. For this article, it would analyse the power status but it would not apply in the L.V. switchboard. Most of the power analyser which sells in Hong Kong is connecting with RS485 Modbus for export the power status to the server or computer.

## 3. METHODOLOGY

### 3.1. Introduction

To measure the power quality in a L.V. switch board, different materials were used. For example, the current transformers, a power analyser, cables etc. In this project, the digital power analyser would be used for measuring a single-phase water pump's power supply quality.

To collect the information of the power analyser, a communication system would be needed. Therefore, after installing the power analyser to the electrical load, the communication system also would be set up with Modbus RS485. The information would be transmitted to the computer and recorded to the database by PLC programme.

To make an Android Smartphone can access the database to monitoring the power system, an Android application would be created. The application can access the database though the internet by Wi-Fi, 4G etc.

The system of this project would be shown in the following Figure 3.1.1.

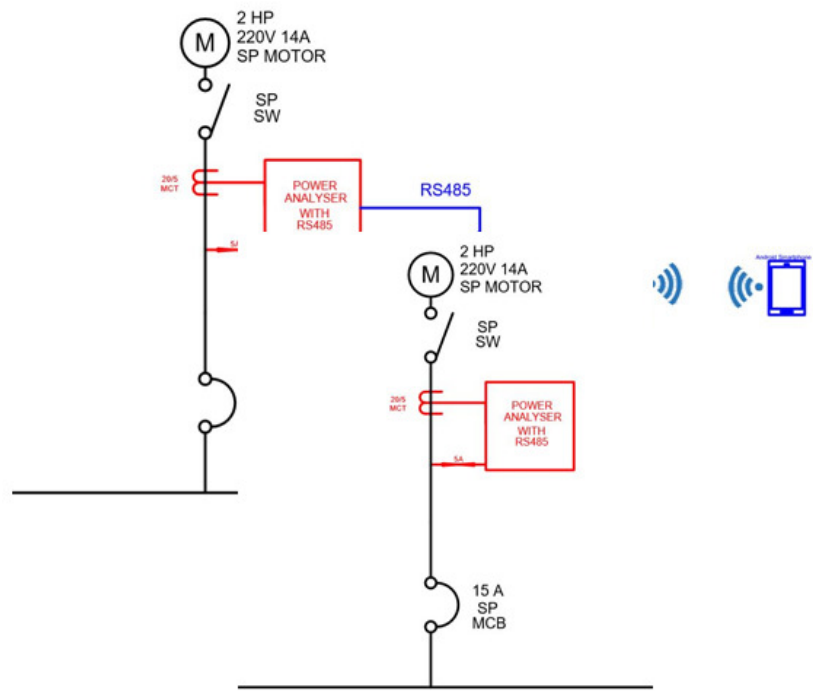

Figure 3.1.1 System of the project

## 3.2. Installation of a Power Analyser

In L.V. switchboard, the power analyser installed after the MCCB or ACB to measure the power status when the MCCB or ACB closed. For the connection of the analyser in this project, which is similar with the connection of the L.V. switchboard, it will show in figure 3.2.1.

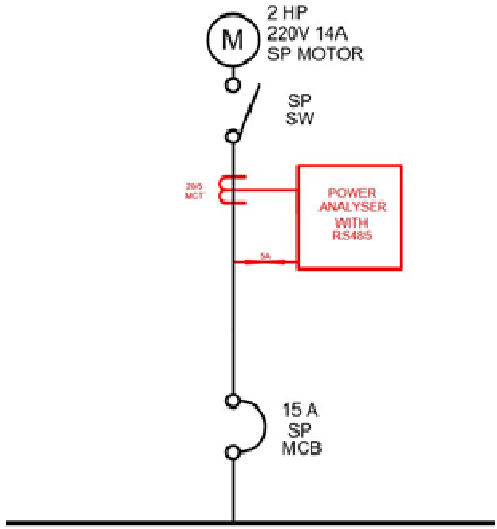

Figure 3.2.1 Schematic diagram for installing power analyser

For the current transformer, it is going to measure the current with the power analyser. There are two types of current transformer are usually use in a L.V. switchboard, the protective current transformer and measuring current transformer. The protective current transformer will be going to carry fault current and link to the IDMT protective relay for protect the circuit with earth fault or overcurrent. The measuring current transformers are carrying full load current for measuring the current with higher accuracy with protection current transformer and connect to the power analyser for reading current. Therefore, a measuring current transformer would be applied in this section.

To measure the voltage, the cable will be connected to the supply directly. In Hong Kong, the voltage would be supplied about 380V line-to-line voltage/ 220V line voltage at the L.V. switchboard from the electrical power supply company.

To protect the power analyser, a 6A fuse would be applied for avoiding the overcurrent fault. The fuse will cut of the circuit to the power analyser when the current is over 6A. After the connection from the L.V. switchboard to the power analyser, the power analyser can measure the voltage, current and frequency, etc.

For the safety, the main incoming switch/circuit breaker should be turn off before installing the power analyser. In this project, the current transformer is already installed in the power analyser. Therefore, the power cables can install to the power analyser directly. To install the power analyser in this project, remove the cable from the switch first. Next, divert the cable to the power analyser. Finally, install new cables to the switch which before the pump.

The connection needs to connect correctly to let the power analyser operate successfully. The monitor of the power analyser is showing the voltage of the supply, it should be 220V plus or minus 6% for a single-phase circuit in Hong Kong according to the supply rule of the electrical power supply company. Therefore, the power analyser is operating normally with a high accuracy monitoring.

### 3.3 Communication to Computer

There are different communication systems to share the information. In most of the power analyser, RS485 Modbus would be applied to communicate to computer and share the information by the request signal. There are different methods to connect the RS485 to the computer. For example, using ARDUINO, an adapter, or connect to a server which designed by the power analyser's manufacturer. A RS485 to USB adapter was applied for connecting the power analyser to the computer in this project.

The Transmission Control Protocol/Internet Protocol (TCP/IP) (Stevens, W.R. [12]) address would be set for different power analyser in L.V. switchboard. In this project, only one power analyser would be used so the IP address would be 1. In the L.V. switchboard, the computer/server can check different power analyser by checking different power analyser with different IP.

To communicate between the computers to the power analysers, the floating-point format data would be applied. The float format is the standard of IEEE-754 [13] which is the interchange and arithmetic formats and methods for decimal and binary floating-point arithmetic in computer programming environment. Said by Sasidharan, M.A. and Nagarajan, M.P.[14]. The floating-point data are usually use on different software and hardware.

### 3.4 Upload Data to the Database from Computer

For recording the data and save to the database, a PLC programme would be used. Programmable Logic Controller (PLC) is a controller which is programmable, can be used to control the sending and receiving signal to the power analyser though RS485 Modbus to the computer. In this project, LabView would be used for inserting data to the database from the computer.

LabView is a software which developed by NI is a graphic programme for measurement and control. The programme would be built by the blocks. The project interface and the block diagram interface would in different window to design. LabView can help to programming the graphic diagram. Different function can be made by the diagram blocks. The blocks connected together then the programme would be done.

In this project, a LabView programme would be developed for receive the data of power status from the power analyser by the computer with the RS485 Modbus connection. The block diagram is shown in the below Figure 3.4.1.

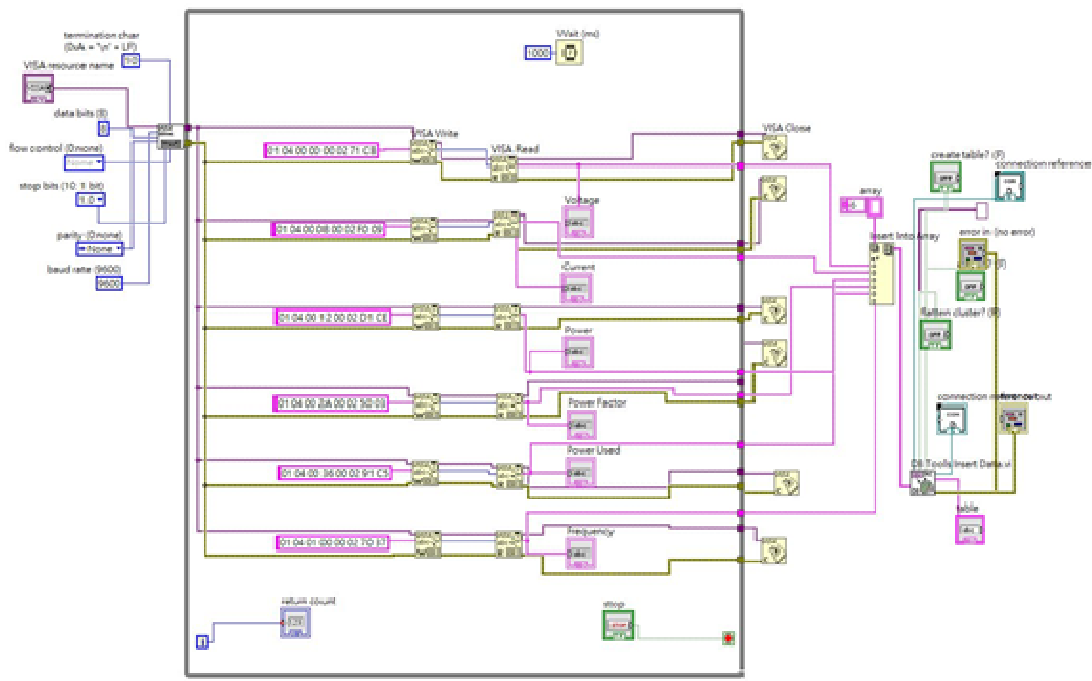

Figure 3.4.1 LabView Graphic programme

For the VISA resource, the setting should be matched to communicate with the power analyser. For this power analyser, baud rate is 9600, parity is 0, the data bit I 8 and the stop bit is 1.0. With select the right port from the computer so the computer can connect to the power analyser successfully. The "For Loop" would be used for taking the new power status. The delay time was set as 1000 mini-second (1second).

According to the user manual, the power analyser would be showing the different power status by input different floating-point format input. For example, input "01 04 00 00 00 02 71 CB" can measuring the voltage for this power analyser. For the first "01" means the IP address of the power analyser. The next "04" are the signal to the power analyser and make a function. The next "00 00" is the function code for reading voltage. The next "00 02" is going to let the power analyser give 2 words of floating-point format number. And the final "71 CB" is the Cyclic Redundancy Check for the previous command. The power analyser in this project can measure the voltage, current, frequency, power factor, the real power and record the power used in kWh.

The results will be show by changing the function code. The programme will keeping write and read because of the "For Loop" function. The Figure 3.4.2 would show how the diagram programme work.

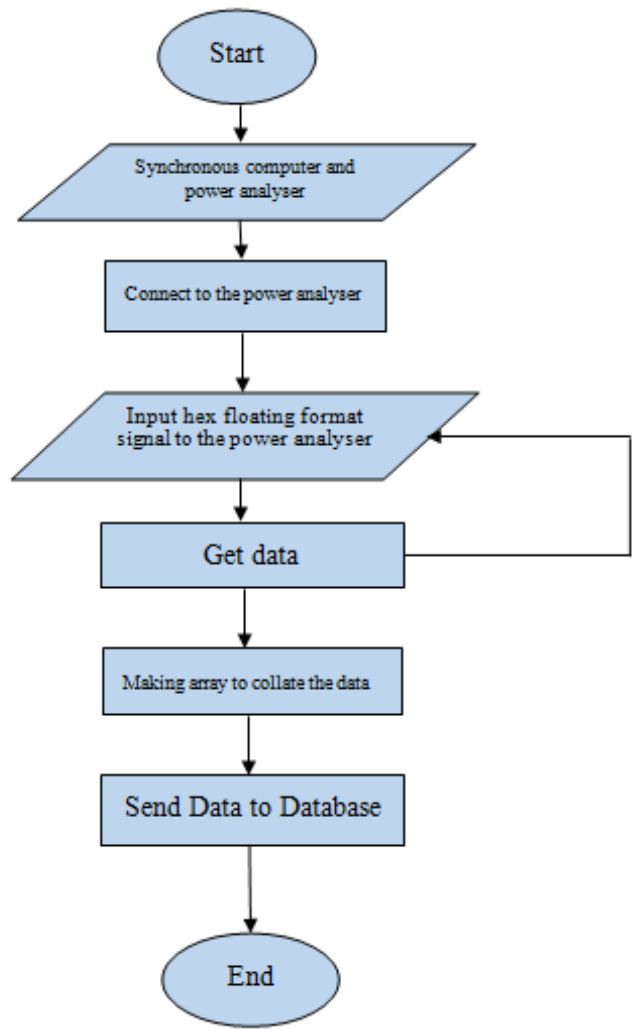

Figure 3.4.2 Flowchart of taking data from the power analyser

After sending the data to the computer, a database should be connected to record the power status. In this project, My SQL would be used to storage the data of the power status. 6 different signals would be sent out from the power analyser, the array function would be used to make up the data to the database.

For the MySQL, a domain server would be used to storage the data. Most of the servers are needed to pay for the service of the cloud server. In this project, a computer would be done as a local host server to be a domain server for record and sending data
.
**3.5 Android Application for Checking Data**

Using an Android device, an Android application would be used. In this project, an android application would be developed for checking the power quality's data by Android studio.
Android is an operating system invented by GOOGLE, said by Lasky, Jack, Salem [15]. It is a popular operating system for smartphones. Most of the smartphone are using android system to operate the applications. It is also an open sourced operation system for different developer to

International Journal of Computer Science & Information Technology (IJCSIT) Vol 11, No 4, August 2019

develop android applications for the electronic device which is using Android operating system. Android Studio is one of the applications for developing an android application.

Android Studio is an application which is using JAVA to write the programme. There are different languages for programming, such as C++, Python and JAVA. Lockhart, Luke E.A. [16] said that Sun Micro systems develop the JAVA cross-platform software in the 1990s and released to programmers in 1996. JAVA also is an open sourced programming that the developer can use this language to develop different application.

To develop the Android application in the Android studio, the user interface would be designed at the .xml file. The components can be chosen in the palette corner. The components need to be set and the setting can be change in the design layout or in the text layout. To make the components get function, the java programme would be used. The programme would be built in the .java file. In the .java file, the components which set in the design interfaces, need to be identified for making function. Different function can be made in the .java file by coding the status.

There different resource can be used from the internet. To add those resources to the Android Studio, the resources need to add in to the "build.gradle" file. Otherwise, the resources cannot be used. After the resource links add into the programme, it should be import in the .java file to use the resource.

For the User Interface, the android application would be access to the activity which using the list view function that recorded the power status from the database. The user can check the power status in anywhere through the internet by the application.

For a power monitoring system, a Login function is needed for the security to let the data not show to other people who has no relationship to the power system. Figure 3.5.1 is showing how the Login function work and the design view. First, the items such as the button and the text input box should be identified for the function. The function "if" would be used in this login function. When the button clicked, the "if" function run. If the input text is equal to my setting in this programme, the application would take the user from the login activity to the power monitoring system activity.

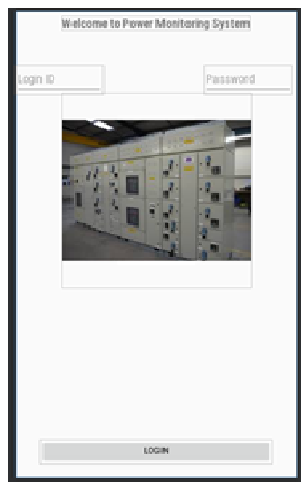

Figure 3.5.1 Design View of the Login function

After the Login activity, the power monitoring system activity would be showed. It has been designed as Figure 3.5.2 and there are different functions in this activity.

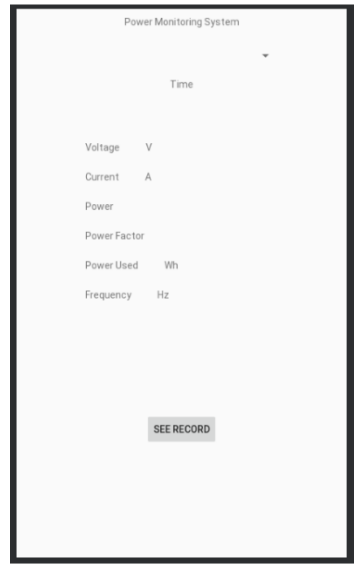

Figure 3.5.2 Design View of the power monitoring system

In this Android Application, the clock would be provided for the user who can know the date and time in the application. It is also can take a screenshot with time when the user wants to record the on-time status. The time would be updated in every second.

For the spinner function, it is going to simulate when the system is going to apply on a power monitor system which is more than one power analyser. The user can choose the power analyser which he wants to check and choose it in the spinner function. Figure 3.5.3 would show the User interface.

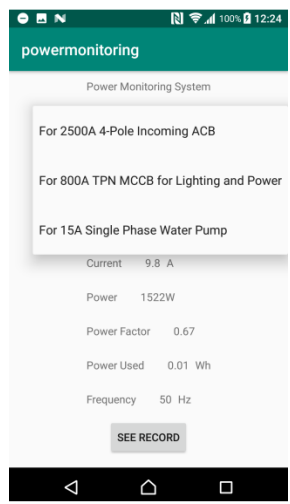

Figure 3.5.3 User Interface of Spinning function

The "switch" function had been used in here. When the target item have been seleted, the function will be work case by case. The textView would be change automatically when the different item selected. In this project, only the 14A Single Phase Water Pump can be recorded in the actrual value.

There is also a function to check the previous record. By using the "SEE RECORD" button, the previous record would be shown in the other activity. The data would be taken from the database with the internet.

To access the database, a PHP file would be used. The PHP file can show the data from the database. Figure 3.5.4 and 3.5.5 was showing how the PHP works.

```php
<?php
define('HOST','localhost');
define('USER','root');
define('PASS','');
define('DB','pm01');

$con=mysqli_connect(HOST, USER, PASS, DB) or die ('ERROR');
?>
```

Figure 5.3.4 Login.php

```php
<?php
include_once("login.php");
$query = "SELECT * FROM `pm01`";

$result = mysqli_query($con, $query);
$number_of_rows = mysqli_num_rows($result);

$response = array();

if($number_of_rows>0){
        while($row = mysqli_fetch_assoc($result)){
                $response[] = $row;
        }
}
header('Content-Type: application/json');
echo json_encode(array("pm01"=>$response));
mysqli_close($con);
?>
```

Figure 3.5.5 PHP file for showing data in JSON

The PHP file is going to login to the database for taking the data because of the security reason.

The JSON file would be export and the database will close when the data took out.

The Android application working would show in a flowchart which showed in Figure 3.5.6.

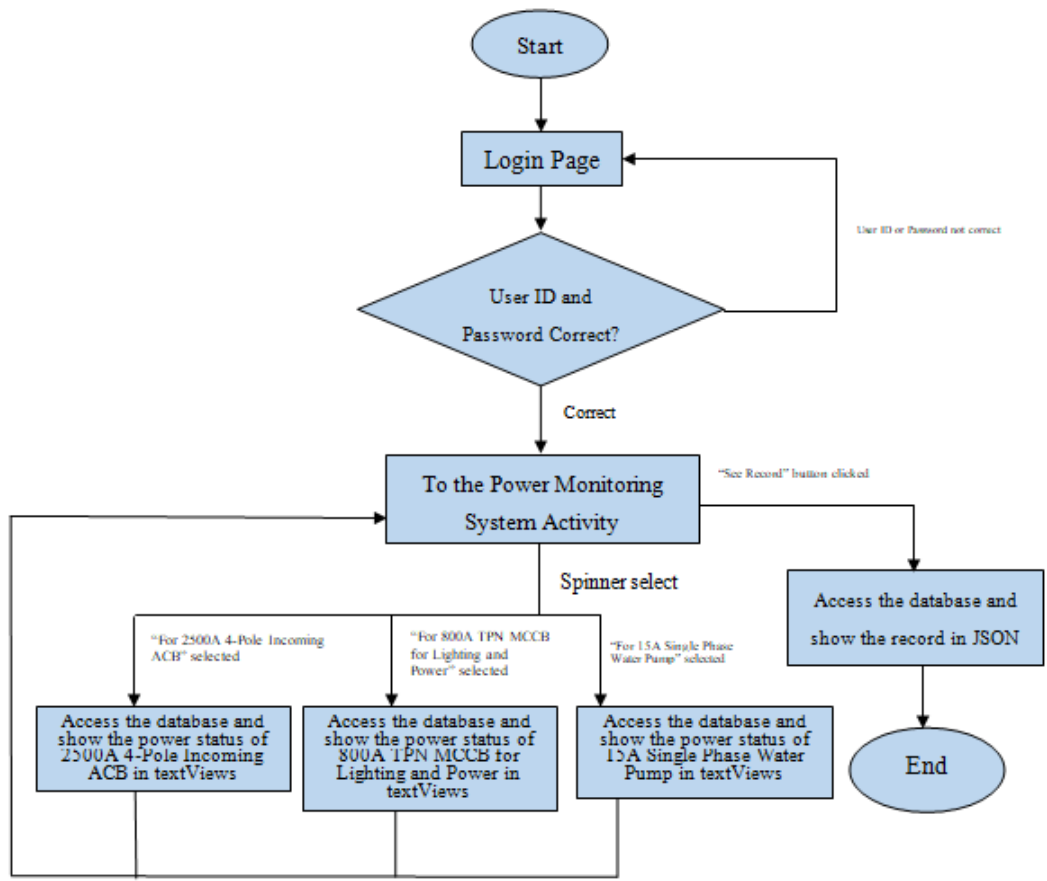

Figure 3.5.6 Flowchart of the Android application.

showed on the power analyser and send the data to the database for record through the RS485 Modbus. An Android application has been developed for showing the real time power status and check the record through the database.

## 3. CONCLUSIONS

In this project, an android smartphone can access the database which recorded the power status from the power analyser for simulating the L.V. switchboard's power monitoring. A single-phase power analyser would be installed to a water pump for simulating the power status would be communicate through a server or computer to the internet for the power monitoring.

For the installation, the new cables have been used for the connection to the pump. The cable sizing also includes in the planning. Because of the pump spends 14A to operate, 1mm2 PVC Copper Cable can be used by according the Code of Practice [17] by the Electrical and Mechanical Service Department of Hong Kong because the 1mm2 PVC copper cable do not have enough current carrying capacity for 14A and the PVC may burn if the current carrying capacity are over. Therefore, the 1.5mm2 PVC copper cable would be used for connecting the power analyser to the electrical load. A RS485 Modbus cable also connected between the computer and the power analyser. A RS485 to USB adaptor would be used for converting the data. For inputting data into the database, a Lab View graphic PLC programme has been made. The programme put the data into the database with the array function.

For the data communication, the data have been taking out by the RS485 Modbus with the computer. The master-slave system has been used as the computer be the master server and the power analyser be the slave device. A LabView graphic programme has been made for taking data from the power analyser though the computer to the database. The data would be shown by different input. The data would take in every second by the loop of the programme. The programme also would synchronous the new data to the database in every second. Most of the power analyser sells in Hong Kong are using the RS485 Modbus in 2wire connection to the server for the L.V. switchboard. Therefore, this project can be simulating for the L.V. switchboard. Although there is some wireless monitoring system provided from the manufacturer of the power analyser, a smartphone application can make more convenience to monitoring the L.V. switchboard power status.

For the data record to the database, the LabView and PHP files would be used. The data recorded from the power analyser will synchronous to the database by the LabView graphic programme.
The PHP files would help to access the database and take the data through the internet. In the L.V. switchboard power monitoring system, a server has been applied for collecting data. By the IP address, the system can classify which the data is from.

For the Android application, a power monitoring system would be developed. It is going to simulation the power monitoring system which is wiring to the computer. The application has a login system that only the people who have the relationship with the power system can access. In this project, three activities have been developed. The "MainActivity" (Login page), "Power_monitoring_system", and the "Database".

The activity for the power monitoring is the "Power monitoirng_system" activity. The components of the activity such as TextView, Spinner etc. should be setup for the user interface. They also need to define in the JAVA programme. The user can monitor the power status in the system through the internet. The record of the power status can be read in the "Database" activity. The activity would access the database through the network and take the data by the PHP command. The JSON file would be took in the listview function at the "Database" activity.

After the power analyser installed and the programmes developed, the power monitoring system created for the 14A single phase water pump electrical load. In the L.V. switchboard power monitoring, more power analysers would be applied and the power monitoring system would help to monitoring the power status.

**AUTHORS**

Tony Tsang (MIEEE'2000) received the BEng degree in Electronics & Electrical Engineering with First Class Honours in U.K., in 1992. He studied the Master Degree in Computation from Oxford University (U.K.) in 1995. He received the Ph.D from the La Trobe University (Australia) in 2000. He was awarded the La Trobe University Post-graduation Scholarship in 1998. Dr. Tsang earned several years of teaching and researching experience in the Department of Computer Science and Computer Engineering, La Trobe University. He works in Hong Kong Polytechnic University as Lecturer since 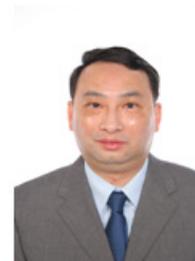 2001. He works in Hong Kong College of Technology in 2014. He has numerous publications (more than 110 articles) in international journals and conferences and is a technical reviewer for several international journals and conferences. His research interests include mobile computing, networking, protocol engineering and formal methods. Dr. Tsang is a member of the IET and the IEEE.